\documentclass[journal, twoside, onecolumn, draftclsnofoot,12pt]{IEEEtran}

\usepackage{multirow}  \usepackage{cite}
\usepackage{graphicx,subfigure}\usepackage{amsmath} \usepackage{amssymb} \usepackage{amsthm}\usepackage{color,array}
\usepackage{etex,etoolbox} \usepackage{float}
\usepackage{color}\usepackage{graphicx}
\usepackage{caption}\usepackage{xcolor}
\usepackage{colortbl}
\usepackage{soul}
\usepackage{algorithm}
\usepackage{soul,color,bm}
\usepackage{setspace}

\usepackage{algorithmic}   \usepackage{bbm}
\usepackage[T1]{fontenc}

\makeatletter
\providecommand{\@fourthoffour}[4]{#4}
\def\fixstatement#1{%
  \AtEndEnvironment{#1}{%
    \xdef\pat@label{\expandafter\expandafter\expandafter
      \@fourthoffour\csname#1\endcsname\space\@currentlabel}}}

\globtoksblk\prooftoks{1000}
\newcounter{proofcount}
\stepcounter{proofcount}
\long\def\proofatend#1\endproofatend{%
  \edef\next{ \Alph{proofcount}. Proof of \pat@label \noexpand\begin{proof}[Proof]}%
  \toks\numexpr\prooftoks+\value{proofcount}\relax=\expandafter{\next#1\end{proof}}
  \stepcounter{proofcount}}

\def\printproofs{%
  \count@=\z@
  \loop
    \the\toks\numexpr\prooftoks+\count@\relax
    \ifnum\count@<\value{proofcount}%
    \advance\count@\@ne
  \repeat}

\makeatother

\fixstatement{thm}
\fixstatement{lem}

\begin{document}


\title{\huge{Deep Cooperative Sensing: Cooperative Spectrum Sensing Based on Convolutional Neural Networks}}
\author{Woongsup Lee,~\IEEEmembership{Member,~IEEE,}  Minhoe Kim,~\IEEEmembership{Student Member,~IEEE,} and Dong-Ho Cho,~\IEEEmembership{Senior Member,~IEEE} 
\thanks{W. Lee is with the Department of Information and Communication Engineering, 
Gyeongsang National University, Republic of Korea. M. Kim and D.H. Cho are with the
School of Electrical Engineering, Korea Advanced
Institute of Science and Technology.}}

\maketitle



\begin{abstract}

In this paper, we investigate cooperative spectrum sensing (CSS) in a
cognitive radio network (CRN) where multiple secondary users (SUs)
cooperate in order to detect a primary user (PU) which possibly
occupies multiple bands simultaneously. Deep cooperative sensing (DCS), 
which constitutes the first CSS framework based on a convolutional 
neural network (CNN), is proposed. In DCS,
instead of the explicit mathematical modeling of CSS which is hard
to compute and also hard to use in practice, the 
strategy for combining the individual sensing results of the SUs
is learned with a CNN using training sensing samples.
Accordingly, an environment-specific CSS which considers both 
spectral and spatial correlation of individual sensing outcomes, 
is found in an adaptive
manner regardless of whether the individual sensing results are quantized
or not. Through simulation, we show that the performance of
CSS can be improved by the proposed DCS with low complexity
even when the number of training samples is moderate. 

\end{abstract}

\vspace{-3mm}
\section{Introduction} \vspace{-0.1cm}

In cognitive radio networks (CRNs), secondary users (SUs) dynamically
utilize the unused channels which are owned by primary user
(PU). Given that the top priority of SUs is not
to interrupt the operation of the PU, it is of utmost importance to determine
whether the PU is present or not. Consequently, spectrum sensing
to detect the presence of the PU is one of the most
important research topic in CRNs \cite{Peh2009, Lee2011,
So2016, Reyes2016, Xue2014, Min2009, Thilina2013}. 

\begin{figure}
\centerline{\includegraphics[width=7.2cm]{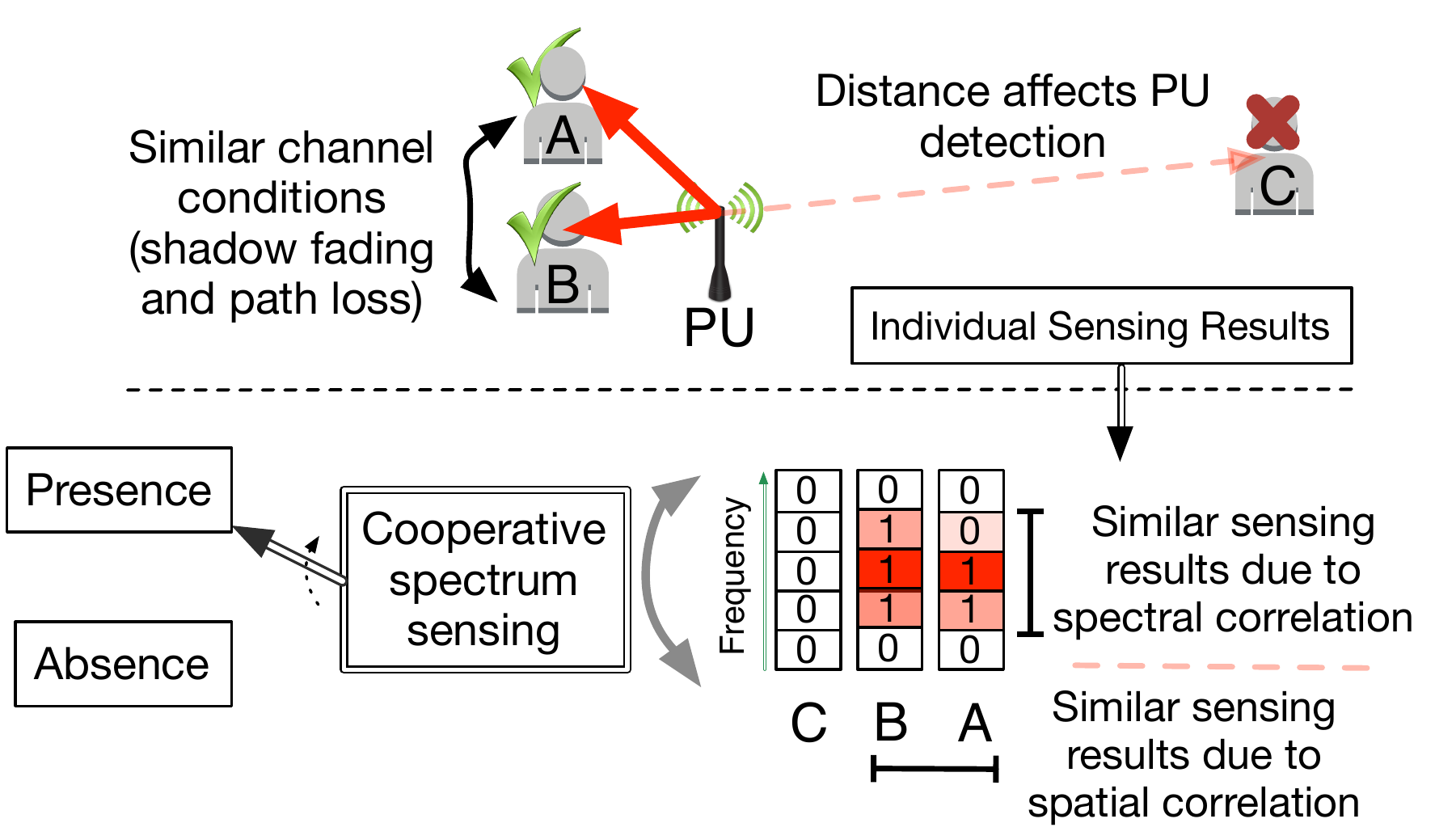}}\vspace{-2mm}
 \caption{CSS with correlated individual spectrum sensing.}
\label{system_model_fig}\vspace{-5mm}
\end{figure}

Given that the sensing results of individual SUs are susceptible to
errors due to the fluctuating channel conditions, cooperative spectrum
sensing (CSS), where the individual sensing
outcomes from multiple SUs are combined to determine the presence
of the PU, has been proposed. However, deriving the
optimal CSS strategy is complicated due to the correlated channel conditions
which depends on the network environment where each SU is situated \cite{Lee2011}.
For example, SUs which are close to the PU (e.g., SU A and 
B in Fig. \ref{system_model_fig})
are likely to detect the PU more reliably than SUs which are
far away from the PU (e.g., SU C in Fig. \ref{system_model_fig}). 
Moreover, because of the spatial correlation of 
wireless channels, SUs which are close to each other (e.g., SU A and B in
Fig. \ref{system_model_fig}) are likely to report similar
sensing results \cite{Xue2014, Min2009}. In addition, the leakage of
transmit power to adjacent bands, i.e., spectral correlation, also needs to be taken into
account. The problem becomes more complicated if the
SUs and PU can change their positions over time.

For this reason, a simplified system model is widely assumed for CSS which entails a performance 
degradation. In particular, simple yet efficient
CSS strategies such as the $K$-out-of-$N$ scheme, which decides that
a PU is present if $K$ SUs out of $N$ SUs detect the PU, are widely considered in the
literature \cite{Peh2009, Lee2011}, where the individual sensing of SUs 
is considered to be independent to each other.
On the other hand, the authors of \cite{Xue2014} and \cite{Min2009} considered the CSS by taking
into account the multidimensional correlation in individual sensing results, 
however, the static SUs without mobility are considered in both works and the proposed scheme 
requires either the location information of SUs \cite{Xue2014} or the distribution of received
signal strength (RSS) of each SU \cite{Min2009}, which are hard to obtain in practice. 

Recently, deep learning has gained considerable attention in many fields
of computer science, especially in image processing, speech
recognition, and natural language processing, due to the significant
performance gains it can achieve compared with conventional 
schemes \cite{Shea2016, Krizhevsky2012, Simonyan2014}.
A deep neural network (DNN) is a multi-layer network which
emulates the neurons of the human brain. In a DNN, the close to optimal
strategy to classify data, e.g., speech and images, is learned
from a large amount of sample data via the back-propagation algorithm,
without the need for developing complicated hand-crafted mathematical models for the data.
By building a sufficiently large network, the DNN based model can
emulate the behavior of highly nonlinear and complicated systems.

Although DNNs have been mainly applied for image processing and
speech recognition so far, they can also be applied in wireless communication
systems, especially for classifying communication signals
\cite{Shea2016, OShea2016a}. Given that the surrounding
environment, e.g., topology of PU and SUs, can be learned, the DNN 
based scheme is more adaptive compared with conventional CSS schemes. 
In \cite{Shea2016}, the
spectrum sensing of a single SU was considered and DNNs were
shown to yield a significant performance gain over conventional schemes.
Moreover, the authors of \cite{OShea2016a} showed that the
data traffic type can be determined accurately with
Long Short Term Memory (LSTM), which is a special type of a recurrent neural 
network (RNN). In addition, the CSS schemes based on other 
machine learning techniques except DNN, e.g., support vector machine (SVM), 
were proposed in \cite{Thilina2013}. 

In this paper, we propose deep cooperative sensing (DCS) which is a new DNN-based
CSS scheme for CRNs. The contributions of this paper can
be summarized as follows.

\begin{enumerate}
\item We propose DCS, which is a CSS scheme for CRNs
employing convolutional neural networks (CNNs), 
a particular type of DNN. In the proposed scheme, the
strategy for combining individual sensing results is learned	
from training samples, independent of the type of sensing 
decisions at individual SUs, i.e., both hard decision (HD) and soft decision 
(SD) sensing are included \cite{So2016}. 

\item DCS exploits the spatial and spectral correlations
of the channels such that the combining strategy for CSS is optimized
according to the location of the SUs and
the characteristics of the PU in an adaptive manner. 
To the best of the authors' knowledge, this is the first
work that applies deep learning for CSS.

\item We evaluate the performance of the proposed scheme based on
computer simulations. Our results confirm that the proposed scheme
can achieve higher sensing accuracy compared with conventional approaches
even when the size of CNN structure is small, i.e., low computational complexity.
Moreover, we show that the proposed scheme achieves sufficiently low
sensing errors with small number of training samples, 
which facilitates the adoption of the proposed scheme in practice. 

\end{enumerate}

The remainder of this paper is organized as follows. In Section II,
we describe the system model. The proposed DCS
scheme is introduced in Section III. Simulation results are provided
in Section IV, and Section V concludes the paper.

\vspace{-0.1cm}
\section{System Model and Spectrum Sensing}\vspace{-0.1cm}

\subsection{System Model}\vspace{-0.1cm}
We consider the interweave CR system in which SUs opportunistically 
utilize the idle bands of PU which are found by the CSS \cite{Lee2011}. 
We assume that $N_{\textrm{SU}}$ SUs and a single PU are randomly
distributed in a given area and move with velocity $v$
such that the locations of the SUs and the PU change over time,
where the locations of SUs are not known to a fusion center which combines
the individual sensing results. 
Moreover, we assume a multi-channel system
with $N_{\textrm{B}}$ bands whose bandwidth is $W$. 
The PU can simultaneously utilize $N_{\textrm{B}_{\textrm{P}}}$ 
consecutive bands, while the SUs do not know which bands are used by the PU.
Furthermore, we assume that the transmit power of the PU in a given band is fixed to
$P$ and can be also leaked to adjacent bands whose proportion of power
leakage is $\eta$. In addition, we consider the additive white Gaussian noise (AWGN) 
whose power spectral density is $N_{0}$, where $w^{j}_{i}(m)$ is the noise
of SU $i$ on band $j$ at time $m$.

A simplified path-loss model with path-loss exponent $\alpha$ and
path-loss constant $\beta$ is adopted for modeling the channel between the
PU and the SUs. Let $d_{i}(m)$ be the distance between the PU and
SU ${i}$ at time $m$, then the path-loss is $\beta (d_{i}(m))^{\alpha}$.
The effect of multi-path fading, $g^{j}_{i}(m)$,
is also considered where $i$ is the index of the SU, $j$ is the
index of the band, and $m$ is the time index. In this work, $g^{j}_{i} (m)$ 
is modeled as an independent
zero-mean circularly symmetric complex Gaussian (CSCG) random variable.

Furthermore, spatially correlated shadow fading is also taken 
into account \cite{Algans2002}. Specifically,
we assume that $h_{i}(m)$ is the shadow fading of the channel between
SU ${i}$ and the PU in dB at time $m$ and it follows a normal distribution
with mean zero and variance $\sigma$ \cite{Algans2002}.
Then, a normal random variable with zero mean and unit variance, 
which we denote as $k_{i}(m)$ (normalized shadow fading), 
can be obtained as $\tfrac{h_{i}(m)}{\sigma}$.
Let SU $\textrm{A}$ and SU $\textrm{B}$ be separated by distance
$d_{A-B}$. Then, the correlation of the normalized shadow fading between these SUs, 
i.e., $k_{\textrm{A}}(m)$ and $k_{\textrm{B}}(m)$, which we denote 
as $\rho_{\mathrm{cor}}(d_{\textrm{A-B}})$, is modeled as \cite{Algans2002, Xue2014, Min2009}\vspace{-1mm}
\begin{equation}
\begin{array}{lll}
\rho_{\mathrm{cor}}(d_{\textrm{A-B}}) =  \mathbb{E} \left[k_{\textrm{A}}(m) k_{\textrm{B}}(m) \right] = e^{\left(-\frac{d_{\textrm{A-B}}}{d_{\mathrm{ref}}}\right)},
\label{eq_corr}
\end{array}
\end{equation}
where $d_{\mathrm{ref}}$ denotes a reference distance whose value depends
on the environment, e.g., rural or urban.  Accordingly, the shadow fading of 
two SUs which are nearby will have a high correlation, i.e., the SUs 
experience  similar shadow fading. In our system model, the RSS of the PU at
two SUs which are close to each other will be similar
due to both path-loss and shadow fading, which results in similar sensing 
outcomes.

\vspace{-1mm}
\subsection{Spectrum Sensing}\vspace{-1mm}

We assume that the SUs perform individual spectrum sensing
which is based on the energy detection in
all $N_{\textrm{B}}$ bands for a time period of $\Delta t$
where $N_{\textrm{ED}}$ channel samples are collected for each sensing attempt. 
Moreover, we assume that SUs do not transmit data during CSS, e.g., 
quiet period in the IEEE 802.22  \cite{Lee2011}, such that 
the individual sensing is not affected by the interference from the
transmission of other SUs.

Regarding the state of the PU, two hypotheses, which are 
$\mathcal{H}_0$ and $\mathcal{H}_1$, are considered, where 
$\mathcal{H}_0$ represents the case where the PU is absence 
while $\mathcal{H}_1$ represents the case where the PU is presence.
Then, the received signal of SU $i$ on band $j$ at time $m$, 
which we denote as $y^{j}_{i}(m)$, can be written as  \vspace{-1mm}
\begin{equation}
\begin{array}{lll}
y^{j}_{i}(m) \\ = \begin{cases}
\kappa_{i}(m)  g^{j}_{i}(m) x(m) + w^{j}_{i}(m),  ~~~~~~{\rm for}  ~ \mathcal{H}_1 ~{\rm and}~ j \in B_{P}\\
\sqrt{\eta} \kappa_{i}(m)g^{j}_{i}(m) x(m) + w^{j}_{i}(m), ~~{\rm for}~   \mathcal{H}_1 ~{\rm and}~ j \in B_{A} \\
w^{j}_{i}(m),~~~~~~~~~~~~~~~~~~~~~~~~~~~~~~~ {\rm otherwise} \\
\end{cases}
\label{asym_sha_rev_1}
\end{array}
\end{equation}
where 
$\kappa_{i}(m) = \sqrt{\frac{P}{\beta (d_{i}(m))^{\alpha} 10^{\frac{h_{i}(m)}{10}}} }$ 
and $x(m)$ is the data transmitted by PU at time $m$ where
$|x(m)| = 1$. Moreover, $B_{P}$, $B_{A}$, and $B_{V}$ are the set of bands which are
occupied by the PU, which are affected by the leaked power of PU 
and which are not affected by the PU, respectively. 

Given that the energy detection is considered, the signal strength of channel 
has to be accumulated. Let $T_{i}^{j}$ be the accumulated RSS of SU $i$ on
band $j$. Then $T_{i}^{j}$ can be written as follows.\vspace{-1mm}
\begin{align}
T_{i}^{j} = \frac{1}{N_{\textrm{ED}}}\sum_{m=1}^{N_{\textrm{ED}}} |y^{j}_{i}(m)|^2.
\end{align} 
It should be noted that $N_{\textrm{B}} \times N_{\textrm{SU}}$ individual sensing outcomes 
are collected by the CRN. We let $\textbf{T}$ be the matrix of the accumulated RSS of 
all SUs.

After calculating $T_{i}^{j}$, SU $i$ can report either the 
measured RSS for each band (SD), i.e., $T_{i}^{j}$, or binary 
sensing results for each band using a sensing threshold $\gamma$ (HD),
which we denote as $\hat{T}_{i}^{j}$.
In the latter case, the SU reports $1$ if $\gamma \leq T_{i}^{j}$ and reports $0$
otherwise. Obviously, the signaling overhead is much larger for the SD compared with 
HD, however, a higher sensing accuracy can be achieved with SD.

Based on the individual sensing results, the fusion center of CRN
determines the presence of the PU by combining the 
individual sensing outcomes. The accuracy of spectrum 
sensing can be denoted by the probability of
false alarm ($P_{\textrm{FA}}$) which is the probability that an SU
falsely detects a PU when no PU is present and the probability of
miss detection ($P_{\textrm{MD}}$) which is the probability
that an SU fails to detect a PU when one is present \cite{Lee2011}.

When SD is used, the optimal fusion rule can be obtained 
by comparing two conditional probabilities which are 
$f_{\textbf{T}}(T_{1}^{1}, T_{1}^{2} \cdots T_{N_{\textrm{SU}}}^{N_{\textrm{B}}}  | \mathcal{H}_0)$ and 
$f_{\textbf{T}}(T_{1}^{1}, T_{1}^{2} \cdots T_{N_{\textrm{SU}}}^{N_{\textrm{B}}}  | \mathcal{H}_1)$,
where $f_{\textbf{T}}(T_{1}^{1}, T_{1}^{2} \cdots T_{N_{\textrm{SU}}}^{N_{\textrm{B}}}  | \cdot)$
is the joint conditional probability density function (PDF) of entire sensing results. 
Although $f_{\textbf{T}}(T_{1}^{1}, T_{1}^{2} \cdots T_{N_{\textrm{SU}}}^{N_{\textrm{B}}}  | \mathcal{H}_0)$ 
can be summarized as $ f(T_{1}^{1}  | \mathcal{H}_0) f(T_{1}^{2}  | \mathcal{H}_0) 
\cdots f(T_{N_{\textrm{SU}}}^{N_{\textrm{B}}}  | \mathcal{H}_0)$ which is easy to calculate, 
where $f(\cdot  | \mathcal{H}_0)$ is the conditional PDF of individual sensing result 
when a PU is absence, $f_{\textbf{T}}(T_{1}^{1}, T_{1}^{2} \cdots T_{N_{\textrm{SU}}}^{N_{\textrm{B}}}  | \mathcal{H}_1)$ is hard to calculate due to spatially and spectrally 
correlated wireless channel. Moreover, 
in order to calculate $f_{\textbf{T}}(T_{1}^{1}, T_{1}^{2} \cdots T_{N_{\textrm{SU}}}^{N_{\textrm{B}}}  | \mathcal{H}_1)$, the distance between SUs, and 
the distance between SUs and PU have to be known, which are hard to obtain in the practical system.
The use of optimal fusion rule becomes more challenging 
as SUs and the PU change their positions due to the mobility.

On the other hand, when HD is used, each SU will report $\hat{T}_{i}^{j}$ which is either 0 or 1 
according to the value of $T_{i}^{j}$. Given that the values of $\hat{T}_{i}^{j}$
are not independent to each other, i.e., $\mathbb{E}(\hat{T}_{i_1}^{j_1}\hat{T}_{i_2}^{j_2})
\neq \mathbb{E}(\hat{T}_{i_1}^{j_1})\mathbb{E}(\hat{T}_{i_2}^{j_2})$, 
all possible outcomes of $\hat{T}_{i}^{j}$, whose cardinality is 
$2^{N_{\textrm{B}} \times N_{\textrm{SU}}}$, have to be examined to derive the optimal
decision policy and this requires huge computational overhead, especially when the number of SUs 
is large. The problem can be simplified
by assuming that $\hat{T}_{i}^{j}$ is independent and identically-distributed (i.i.d.)
such that the number of SUs which 
report 0 and 1 is sufficient to derive the final CSS decision\footnote{In this case, 
the number of SUs which report $\hat{T}_{i}^{j} = 1$ when the PU is absence 
($\mathcal{H}_0$) and those report $\hat{T}_{i}^{j} = 0$ when the PU is 
presence ($\mathcal{H}_1$) can be modeled as the Binomial random variable
whose success probabilities are $\textrm{Pr}(\hat{T}_{i}^{j}=1 | \mathcal{H}_0)$
and $\textrm{Pr}(\hat{T}_{i}^{j}=0 | \mathcal{H}_1)$, respectively, and
appropriate decision criteria can be found consequently \cite{Peh2009}.}.  
However, in practice, the effects of the spectral and spatial correlation 
of wireless channel on individual sensing outcomes
have to be taken into account in order 
to achieve higher sensing accuracy. Therefore, from the discussions above, 
we can conclude that the derivation and utilization of mathematically optimal 
CSS is hard in practice and new yet efficient approach which considers
correlations of wireless channel has to be devised.

\vspace{-1mm}
\section{Deep Cooperative Sensing}\vspace{-1mm}

In DCS, the CNN model is used to combine individual sensing results
to determine the presence of the PU. CNNs have been widely used in
deep learning research to classify images by exploiting their
spatial characteristics. In CNN, multi-dimensional convolution
is applied to an image to extract its spatial features. 
The recent success of CNNs in visual classification
\cite{Krizhevsky2012, Simonyan2014} motivates the adoption of CNNs for CSS,
because similar to the correlation of nearby pixels in images, 
the individual sensing outcomes from nearby SUs and adjacent
bands are likely similar due to their spatial and spectral correlation, 
i.e., $\mathbb{E}\left[ (T_{i}^{j} - \mathbb{E}[T_{i}^{j}])(T_{i+e_1}^{j+e_2} - \mathbb{E}[T_{i+e_1}^{j+e_2}])\right]   > 0$ for $e_1, e_2 = -1, 0, +1$.
\begin{figure}
\centerline{\includegraphics[width=5.7cm]{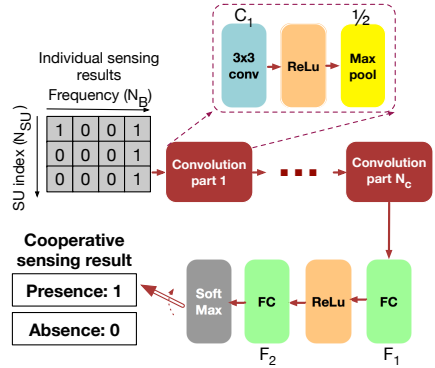}}\vspace{-2mm}
 \caption{CNN model for deep cooperative sensing.}
\label{CNN_model_fig}\vspace{-5mm}
\end{figure}
In the following, we first describe the structure of the DCS and then   
discuss how DCS can be trained to detect the presence of the PU.

\vspace{-2mm}
\subsection{Structure of DCS}\vspace{-1mm}

The proposed CNN structure for DCS is composed of two parts, which are the convolution part 
at the front of the network and the fully connected (FC) part at the back of the network,
as shown in Fig. \ref{CNN_model_fig}. First, in our proposed scheme, individual sensing 
results from different bands and SUs constitute the two dimensional input data 
for the CNN which is fed into the convolution part, unlike the DNN based spectrum 
sensing considered in \cite{Shea2016} where the in-phase and 
quadrature-phase of the temporal signal are used 
to define a two dimensional matrix. The elements of input 
matrix can be either binary for HD or continuous for SD.

In our CNN structure, $N_{\textrm{C}}$ convolution parts which are
composed 3$\times$3 convolution layer (3$\times$3 conv.), rectifier linear unit (ReLU) layer, 
and max pooling layer, are connected in tandem. The convolution layer 
performs spatial convolution of the input data such that the spatial features 
of the input data can be extracted. The size of each spatial filter in the convolution 
layer is set to 3$\times$3 because as can be concluded from 
\cite{Simonyan2014}, the 3$\times$3 convolution is sufficient to extract 
the spatial features of the input data. Accordingly, let $X_{\textrm{conv}}$, $W_{\textrm{conv}}$, 
and $Y_{\textrm{conv}}$ be the input, weights, 
and output of the convolution layer. Then, $Y_{\textrm{conv}}[m, n] = \sum_{\epsilon_1=0}^{2} 
\sum_{\epsilon_2=0}^{2} X_{\textrm{conv}}[\epsilon_1+m-1, \epsilon_2+n-1] 
W_{\textrm{conv}}[\epsilon_1, \epsilon_2]$. 
The depth of the convolution layer for $i$-th convolution part 
is set to $C_{i}$ which can be adjusted according to the size
of input data, e.g., $C_{i}$ has to be large\footnote{However, through simulations, 
we show that a relatively small $C_{i}$, i.e., small sized CNN structure,
compared with what is normally used for image 
classification\cite{Krizhevsky2012, Simonyan2014}, can achieve 
sufficiently high sensing accuracy, because
unlike image classification where typically hundreds of different classes have
to be distinguished, in our system model, we only need to classify two 
classes, namely the presence and absence of the PU.} when $N_{\textrm{SU}}$ and 
$N_{\textrm{B}}$ is large.  
The stride, which is the step size used in the convolution filter, is set to 1 
and zero padding is used, so that the size of the output remains the same as that of the input.

After the spatial features have been extracted, the ReLU layer 
introduces non-linearity to the CNN \cite{Simonyan2014}. 
When the input of the ReLU layer is $X_{\textrm{ReLU}}$, 
the output is given by $\textrm{max}(X_{\textrm{ReLU}}, 0)$, such that negative
inputs are blocked. Without the ReLU, 
the CNN model would be linear and unable to classify
non-linear behavior. The ReLU layer is followed by the max pooling
layer which reduces the size of the data. With the max pooling layer,
the computational overhead can be efficiently reduced without
significant performance loss as shown in \cite{Krizhevsky2012}.

The FC part performs classification based on the output of the convolution part by 
gathering the results of feature extraction. 
To this end, the FC layer mixes all the data from the convolution part using 
the multiplication of the weights and the addition of biases. Specifically, 
the output of $m$-th FC layer, $Y^{m}_{\textrm{FC}}$, becomes 
$Y^{m}_{\textrm{FC}} = W^{m}_{\textrm{FC}}X^{m}_{\textrm{FC}}+b^{m}_{\textrm{FC}}$,
where $X^{m}_{\textrm{FC}}$, $W^{m}_{\textrm{FC}}$ and $b^{m}_{\textrm{FC}}$ are the inputs, 
weights and biases for the $m$-th FC layer, respectively. In DCS, we consider 
two FC layers whose depth (number of hidden node) are $F_{1}$ and $F_{2}$, 
respectively. Moreover, an additional ReLU layer is placed 
between two FC layers in order to introduce non-linearity. 
Then the output of the FC layers is fed into the softmax operator in which the final decision on 
whether the PU is present or not is made using the softmax function. 
Let index $0$ and $1$ denote the absence
and the presence of the PU, respectively, and $X_{\textrm{S}}$ be the input of the
softmax operation. Then, in our model, the CRN determines that the channel is idle if
$\frac{e^{W^{0}_{\textrm{S}}X_{\textrm{S}}}}{\sum_{i=0}^{1} e^{W^{i}_{\textrm{S}} X_{\textrm{S}}} }
> \frac{e^{W^{1}_{\textrm{S}}X_{\textrm{S}}}}{\sum_{i=0}^{1} e^{W^{i}_{\textrm{S}}X_{\textrm{S}}} }$, 
and otherwise it concludes that the channel is occupied by the PU, 
where $W^{i}_{\textrm{S}}$ is a weight matrix of the softmax operator.

\vspace{-2mm}
\subsection{Training of PRNet}\vspace{-1mm}
		
In order to train the proposed DCS,  individual sensing outcomes 
are collected first. Then, the proposed DCS is trained,
i.e., the weights and biases of each layer are found, to minimize the cross entropy loss,
$\mathcal{L}$, using off-the-shelf stochastic gradient descent 
algorithms, i.e., adaptive moment  estimation. \vspace{1mm} Specifically, the cross entropy loss 
can be written as $\mathcal{L} = -\sum^{1}_{\epsilon=0} l_\epsilon \log(Y^\epsilon_{\textrm{S}})$
where $Y^\epsilon_{\textrm{S}}$ is the output of softmax operator,
i.e., $Y^\epsilon_{\textrm{S}} = \frac{e^{W^{\epsilon}_{\textrm{S}}X_{\textrm{S}}}}{\sum_{i=0}^{1} e^{W^{i}_{\textrm{S}} X_{\textrm{S}}} } \vspace{0.1cm}$, and $l_\epsilon$ is the target label whose
value is either 0 or 1. When the PU is presence, $l_0 = 0$ and $l_1 = 1$, and
if the PU is absence, $l_0 = 1$ and $l_1 = 0$. Note that the cross entropy loss is 
minimized when the estimated value matches with the correct label, i.e. 
$Y^\epsilon_{\textrm{S}} = 1$ for $l_\epsilon = 1$, such that the DCS can be
trained to accurately detect the presence of the PU.

In our proposed scheme, the multiple CNN models with different permutations of SU index
are trained simultaneously and the trained model 
whose accuracy is highest, is chosen. Through the permutation, we can find 
the proper order of SU index such that the sensing results of 
neighboring SUs can be placed nearby in the input data, which improves the performance 
of DCS. It is worth noting that the optimal permutation of SU index cannot be known in advance
because the location information of SUs is unavailable. Although we do not show in the performance 
evaluation, we have observed that the sensing error can be decreased by 17\% through the 
permutation of SU index.

Thereafter, the trained model can be used to determine the presence of PU
based on the individual sensing results. Although the training of DCS might require 
huge overhead, the inference of final sensing result can be conducted with 
relatively low overhead as shown later in the performance evaluation such that
the real-time operation of our proposed scheme is feasible.

\vspace{-2mm}
\section{Performance Evaluation}\vspace{-1mm}

In this section, the performance of the proposed DCS is examined.
Our performance evaluation is implemented using Tensorflow which is
an open-source software library for machine intelligence developed by Google. 
We assume that a single PU and multiple SUs which move at $v = $3 km/h,
are randomly deployed in a 200 m $\times$ 200 m area. Moreover,
the number of bands is set to 16 where $W$ = 10 MHz
and $N_{\textrm{B}_{\textrm{P}}}$ is randomly
chosen from 1 to 3, i.e., the PU can utilize upto three bands simultaneously.
Furthermore, we assume that $\eta$ = -20 dB, $P$ = 23 dBm,
$\beta = 10^{3.453}$, $\alpha = 3.8$, $\sigma$ = 7.9 dB, and
$d_{\mathrm{ref}} = $ 50 m \cite{802_202007, Algans2002}. Moreover, 
we assume that $N_{\textrm{C}} = 3$, $C_{1} = C_{2} = C_{3} =
 F_{1} = F_{2} = 8$. Note that we have used a relatively small number 
of weights and biases\footnote{When the number of SUs is 32, 
the number of weights in our CNN model is 2264, which is smaller than
that for image classification (61,000,000) \cite{Krizhevsky2012}.} for CNN 
structure such that the training and inference of DCS can be performed with low complexity.

We assume that each SU performs individual spectrum sensing every
2 seconds, i.e., $\Delta t$ = 2 seconds \cite{Lee2011}.
Threshold $\gamma$ is set to -107 dBm  \cite{Lee2011} such that when
HD is used, each SU sends $1$ if the RSS 
exceeds $\gamma$ and sends $0$ otherwise.
When SD is used, each SU reports the measured RSS, i.e., $T_{i}^{j}$. 
For performance evaluation, $N_{\textrm{sample}}$ samples are used to train the 
model and the 2000 samples are used for the evaluation where these 
data sets are generated by computer simulation\footnote{In this work, a synthetic data set
is used for learning, whereas in a practical system, real data 
would be used. Nevertheless, the performance
evaluation based on the synthetic data can also provide meaningful insights
regarding the performance of the proposed scheme \cite{Shea2016}.}.


For the performance metric, we consider the sensing error which is the sum of 
the probability of false alarm ($P_{\textrm{FA}}$) and the probability of
miss detection ($P_{\textrm{MD}}$) \cite{Xue2014}. We
examine the sensing error of our proposed scheme for both SD and
HD. Moreover, conventional CSS based on the $K$-out-of-$N$ scheme 
is considered where the value of $K$ is selected such that
$(P_{\textrm{FA}} + P_{\textrm{MD}})$ is minimized. Furthermore, conventional CSS
based on SVM with the linear kernel is also considered which shows the
lowest sensing error in \cite{Thilina2013}.

\begin{figure}
\centerline{\includegraphics[width=5.0cm]{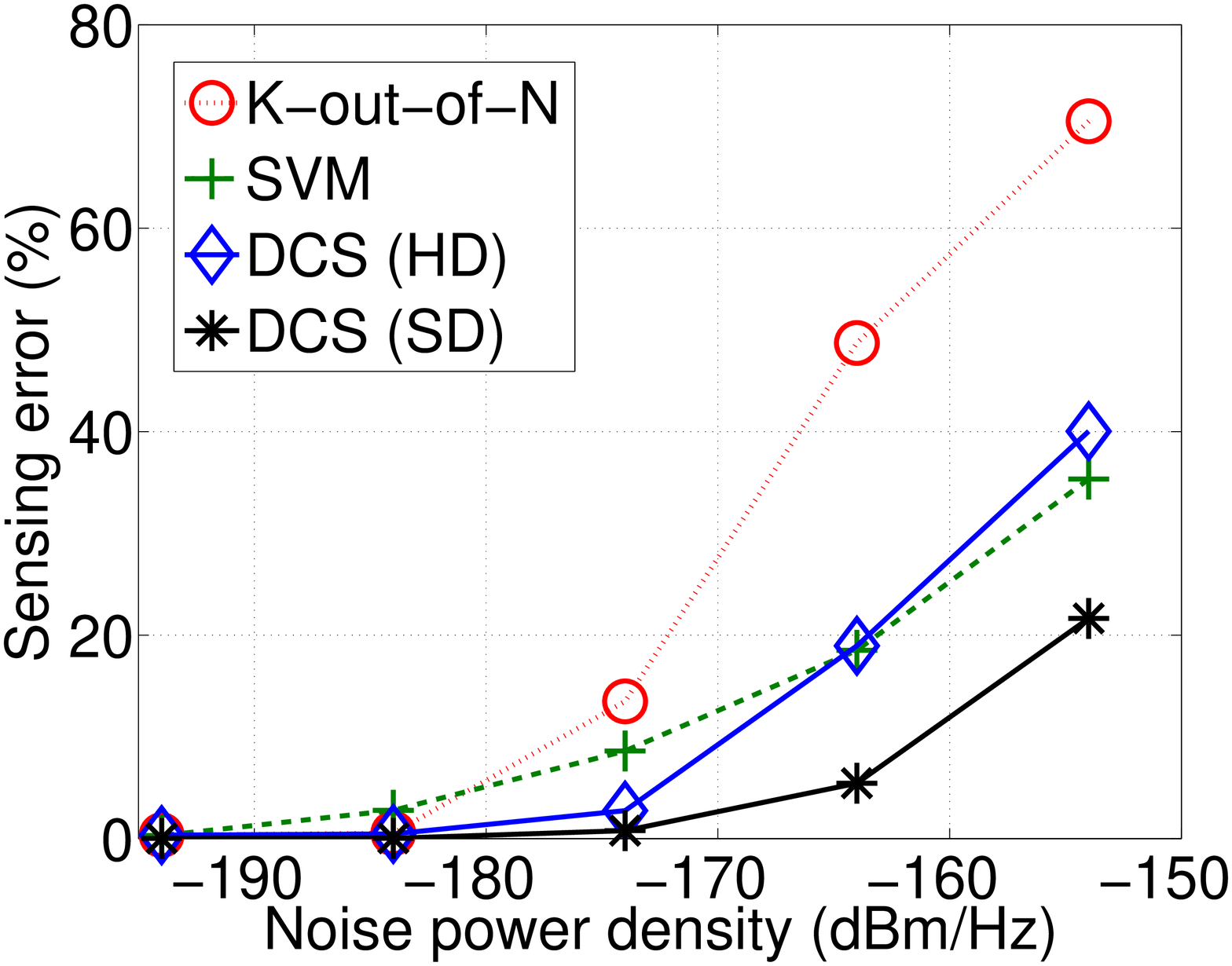}}
 \caption{Noise power density vs. sensing error when $N_{\textrm{SU}}$ = 32 and $N_{\textrm{sample}}$ = 200.}\vspace{-2mm}
\label{fig_noise}\vspace{-3mm}
\end{figure}
\begin{figure}
\centerline{\includegraphics[width=5.0cm]{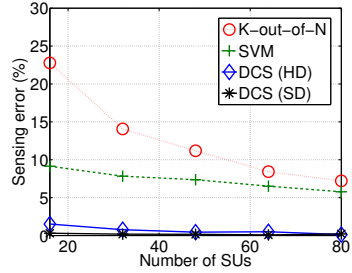}}
 \caption{Number of SUs vs. sensing error when $N_{0}$ = -174 dBm/Hz and $N_{\textrm{sample}}$ = 200.}\vspace{-2mm}
\label{fig_sus}\vspace{-2mm}
\end{figure}
In Fig. \ref{fig_noise}, we show the sensing error vs. the noise density for
$N_{\textrm{SU}}$ = 32 and $N_{\textrm{sample}}$ = 200. We observe that the sensing error
increases as the noise density increases due to the inaccurate individual sensing results.
Moreover, we can also find that the sensing error of DCS is lower than 
those of conventional schemes, i.e., $K$-out-of-$N$ scheme and SVM scheme, when
$N_{0} \leq$ -164 dBm/Hz, and the DCS with SD shows the lowest sensing error in
all cases, which highlights the benefits of the proposed scheme. Especially, we can 
find that the DCS with SD shows sufficiently low sensing error, i.e., sensing error is less than 
21\% even when the power of the noise is very large, e.g., $N_{0}$ = -154 dBm/Hz.
The DCS with HD shows worse sensing accuracy than that with SD because
individual sensing results contain less information \cite{So2016}.

In Fig. \ref{fig_sus}, we examine the sensing error as function of the number of SUs for
$N_{0}$ = -174 dBm/Hz and $N_{\textrm{sample}}$ = 200. As can be observed from 
Fig. \ref{fig_sus}, the performance of CSS is improved as the number of cooperating SUs 
increases, which coincides with our intuition. 
We can also confirm that DCS significantly outperforms conventional schemes,
especially when the number of SUs is small.

\begin{figure}
\centerline{\includegraphics[width=5.0cm]{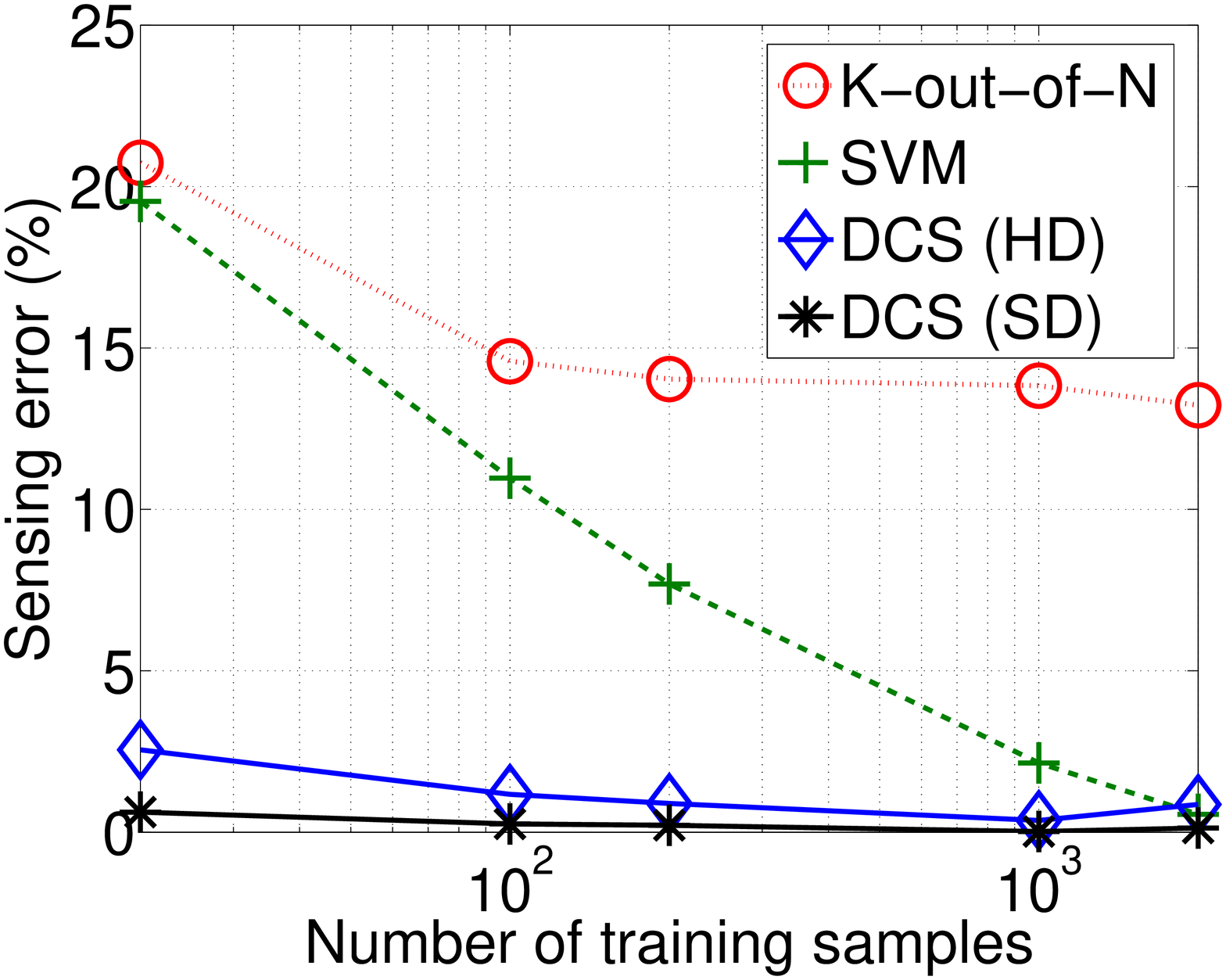}}
 \caption{Number of training samples vs. sensing error when $N_{\textrm{SU}}$ = 32 and $N_{0}$ = -174 dBm/Hz.}\vspace{-2mm}
\label{fig_sample}\vspace{-3mm}
\end{figure}

The sensing error by varying the number of 
training samples for $N_{\textrm{SU}}$ = 32 and $N_{0}$ = -174 dBm/Hz
is shown in Fig. \ref{fig_sample}.
Given that the decision policy of $K$-out-of-$N$ scheme and 
SVM scheme is also determined based on training samples, the sensing error
of those schemes also changes as $N_{\textrm{sample}}$ changes.
We can again confirm that DCS achieves lower sensing error compared with 
conventional schemes. It is worth noting that DCS shows sufficiently low
sensing error even when the number of training samples is small, 
i.e., $N_{\textrm{sample}}$ = 100,
such that the overhead of data collection for training data will be small. 

\begin{figure}
\centerline{\includegraphics[width=5.0cm]{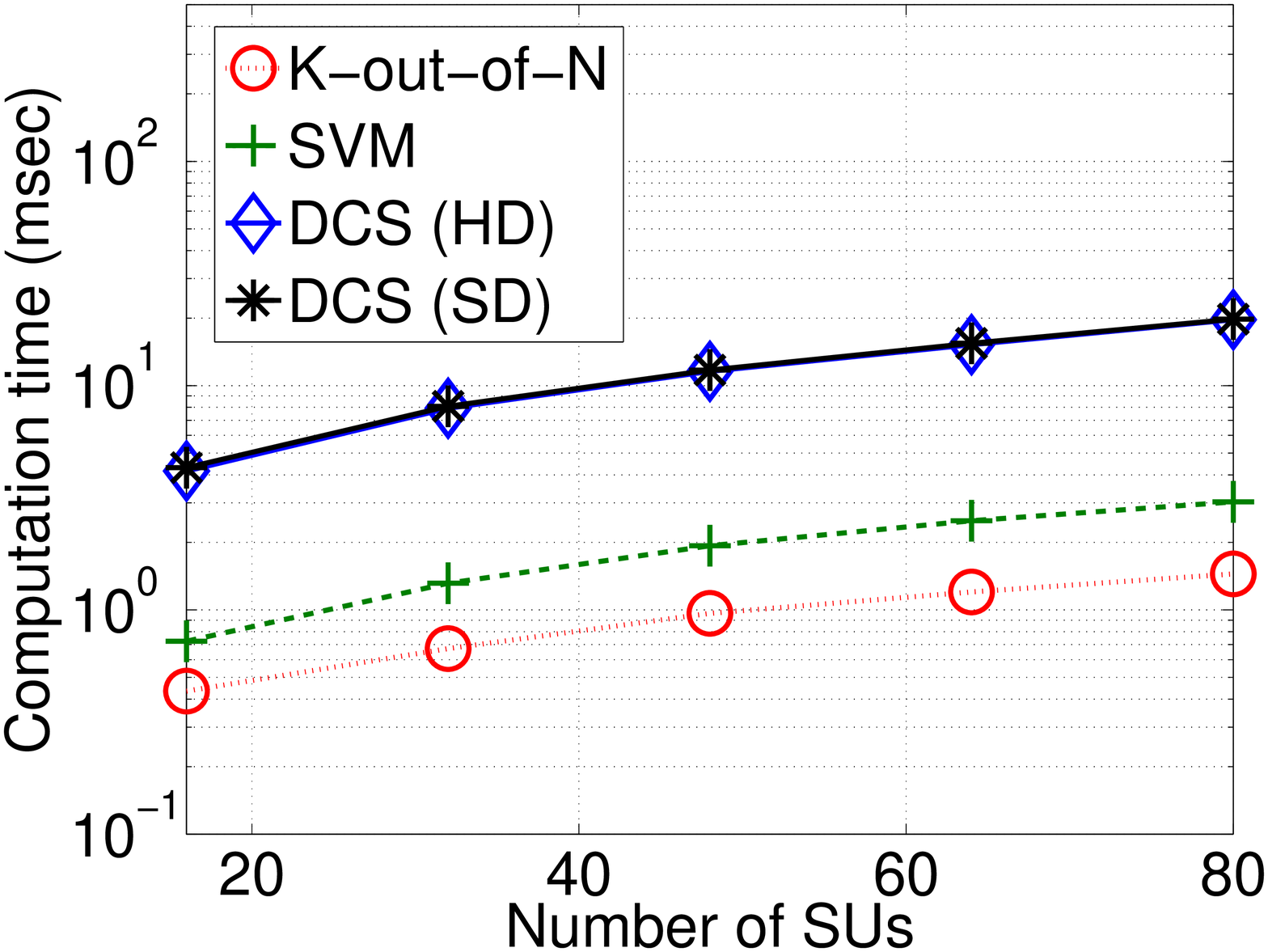}}
 \caption{Number of SUs vs. computation time when $N_{0}$ = -174 dBm/Hz and $N_{\textrm{sample}}$ = 200.}\vspace{-2mm}
\label{fig_comp}\vspace{-2mm}
\end{figure}

Finally, in Fig. \ref{fig_comp}, we evaluate the computation time of considered CSS schemes as function of the number of 
SUs for $N_{0}$ = -174 dBm/Hz and $N_{\textrm{sample}}$ = 200. Although the computation
time of DCS is larger than those of conventional schemes, it is reasonably small, e.g., 
the computation time of DCS is less than 20 msec in all cases, such that 
our proposed scheme can be operated in real-time manner.

\vspace{-1mm}
\section{Conclusions}\vspace{-1mm}

In this paper, a novel CNN-based CSS scheme for CRN 
was proposed, which is the first attempt to use deep 
learning for CSS. In DCS, the strategy for combining the binary or
real valued individual sensing results of the SUs is learned using CNN.
Through simulations we investigated the performance of the proposed 
scheme and found that DCS outperforms conventional CSS schemes. 
We also found that DCS can detect the PU accurately 
with small sized CNN structure, even when the number of
training samples is small, which verifies the applicability of our
proposed scheme in practice.

\vspace{-1mm}
\bibliographystyle{IEEEtran}\vspace{-1mm}
\bibliography{IEEEabrv,mybibfile}

\end{document}